\documentclass[preprint]{aastex}

\shorttitle{Structure of 2MASS edge-on galaxies}
\shortauthors{Bizyaev \& Mitronova}

\begin{document}

\title{Structural Parameters of Stellar Disks from 2MASS Images
of Edge-on Galaxies}
 
\author{Dmitry Bizyaev \altaffilmark{1,2} and Sofia Mitronova \altaffilmark{2}}

\altaffiltext{1}{New Mexico State University and Apache Point Observatory
Sunspot, NM, 88349, USA}

\altaffiltext{2}{Sternberg Astronomical Institute, Moscow, 119992, Russia}

\altaffiltext{3}{Special Astrophysical Observatory, Nizhniy Arkhyz, Russia}

\begin{abstract}
We present results of an analysis of
the J, H, and K$_s$ 2MASS images of 139 spiral edge-on galaxies
selected from the Revised Flat Galaxies Catalog. 
The basic structural parameters
scalelength ($h$), scaleheight ($z_0$), and central surface brightness 
of the stellar disks ($\mu_0$) are determined for all selected
galaxies in the NIR bands. The mean relative ratios of the scaleheights 
of the thin stellar disks in the J:H:K$_s$ bands are 1.16:1.08:1.00, 
respectively.
Comparing the scaleheights obtained from the NIR bands for the same
objects, we estimate the scaleheights of the thin stellar disks
corrected for the internal extinction.
We find that the extinction-corrected scaleheight is, on average, 11\%
smaller than that in the K-band. Using the extinction-corrected structural
parameters, we find that the dark-to-luminous mass ratio is, on average, 1.3
for the galaxies in our sample within the framework of 
a simplified galactic model. 
The relative thicknesses of the stellar disks $z_0/h$ correlates with their
face-on central surface brightnesses obtained from the 2MASS images. 
We also find that the scaleheight of the stellar disks shows no systematic
growth with radius in most of our galaxies.
\end{abstract}


\section{Introduction}

Study of the parameters of galactic components via decomposition 
of the rotation curves is often ambiguous
in the case of spiral galaxies.
The same rotation curve often may be 
explained by models with very different
parameters of the components. The dark-to-light mass ratio may vary 
several times between the models explaining well the same rotation curve,
especially if the latter is smooth and featureless 
\citep[see for example][]{deblok01,bz02,dutton05}. 
Incorporating information about the stellar velocity dispersion 
from observations helps to significantly constrain the masses of galactic 
components \citep{bottema93,khoperskov01}  
and relieve some ambiguities that appear in the modeling of rotation curves,
but observing the distributions of the velocity dispersion 
is a rather difficult task. 
The stellar disk thickness can be used as an alternative to the 
stellar velocity dispersion \citep{zasov02}. 
Studies of the vertical structure of stellar disks alone allow 
the constraint of some dark halo 
parameters \citep{bahcall84,zasov91,zasov02,b03,bk04}.
Note that not all galaxies are suitable for such studies:
inferring the structural parameters for a spiral galaxy with
a significant bulge from photometric data alone is dangerous 
and should be supplemented with concerns of different mass-to-light ratio
for different galactic subsystems \citep[see discussion by ][regarding
difference between photometrically and dynamically identified 
subsystems]{abadi03}. 

Infrared observations are crucial for studies of the structure of edge-on
galaxies. The dust attenuation makes the extraction of the structural
parameters difficult from the optical data. In the NIR bands the galaxies
are much more transparent, which facilitates studies of the structure of 
galaxies. 
The all-sky 2MASS survey \citep{2mass} offers a good opportunity to increase
the number of edge-on galaxies available for studies in the near-infrared
bands. Although faint parts of the galaxies are not seen in the 2MASS images,
the thin disks of the galaxies are obtained with sufficient signal-to-noise
ratio (S/N hereafter).

In this paper we focus on the determination of stellar disk structural
parameters corrected for the internal extinction. One 
approach to the problem is to perform the 
modeling of the radiation transfer in 3-D stellar-gaseous disks 
\citep[like in][]{xi99,kylafis01,bianchi07}. Such studies require 
rather deep images, although there are attempts to apply
this kind of analysis to larger sample of moderate quality \citep{b07}.
In the present paper we develop another approach to study the
extinction-corrected structure of the stellar galactic disks 
comparing the parameters of the disks obtained from different NIR bands. 

In Section 2 we discuss the sample selection and 
evaluation of the stellar disk's structural parameters, 
consider reliability of the parameters, and
study the radial variation of the disk scaleheight.
In Sections 3 and 4 we show how
the comparison of the parameters obtained in the three NIR bands can be used
to estimate the extinction-corrected scaleheight in the galactic disks. 
Section 5 considers the connection between 
the galactic disk thickness with parameters of the dark
halos around the spiral galaxies in the frames of a simplified model. 
Section 6 summarizes results of the paper. 

\section{Sample of Galaxies and Analysis of the Images}

As a pre-defined sample of edge-on galaxies, the Revised Flat Galaxies 
Catalog \citep[RFGC hereafter, ][]{rfgc}
is chosen. The catalog contains visually-classified extragalactic
objects with major-to-minor axes ratio a/b $>$ 7. To secure enough pixels
for studying the vertical structure, we select objects with the major axis
greater than 1 arcmin. This size is estimated directly from the 2MASS   
images at the level of S/N $\sim$ 3. Note that this size is significantly less 
than commonly used diameter D$_{25}$. As a result, we select 139 spiral 
edge-on galaxies that have images in all three 2MASS bands: H, J, and K$_s$.

We apply the technique described by \citet{bm02} to obtain the structural 
parameters of stellar disks of the galaxies disks. This method is based on
the analysis of
photometric profiles drawn parallel to the major and minor axes of galaxy at
a one-pixel interval. The volume brightness in the stellar disks is assumed
to change in the radial $r$ or vertical $z$ directions as follows: 
\begin{equation}
\label{vol_disk}
\rho_L(r,z) = \rho_{L0} exp(-r/h)\, sech^2(z/z_0),
\end{equation}
where $h$ and $z_0$
are the scalelength and scaleheight of the disk, respectively, 
and $I_0 = \int \rho(0,z) dz$ is the central surface luminosity of the
face-on disk, which corresponds to the central surface brightness $\mu_0$.
The model photometric profiles are obtained by the integration 
of equation (\ref{vol_disk}) along the line-of-sight and then by convolution
with the instrumental profile. The best-fit parameters $h$, $z_0$, and $I_0$
are estimated for each radial or vertical profile, and then their 
median values are taken as the resulting scales and central surface brightness
of the stellar disks, whereas the standard deviations of $h$, $z_0$, and $I_0$
estimated from different profiles are taken as uncertainty of the 
parameters. The central regions of the 
galaxies in which bulges may exist are excluded from our analysis.
The parameters $h$, $z_0$, and $\mu_0$ are obtained in the J, H, and K$_s$ 
bands for selected 139 galaxies. The $K_s$-based parameters are shown 
in Table~1. 

Figure~1 shows histograms of distribution of the scale heights (top panel) 
and scale lengths (bottom panel) estimated for our program galaxies. In both
panels, the J- and H-band scales are normalized by those in the K$_s$ band. 
It can be noticed that the scaleheights look systematically larger, on
average, in the J band than those in the K$_s$ for the same galaxy. We will
discuss this effect below: in \S~\ref{three_band} using our observational 
data, and in \S~\ref{im_sim} with the help of artificial data.


\subsection{Radial Variation of the Disk Scale Height}

As in \citet{bm02}, we investigate how the scaleheight $z_0$ of the 
stellar disks in our galaxies change with the distance to the center $r$. 
Our approach to the estimation of the $z_0$ is independent of 
the apriori assumptions about behavior of the scaleheight with radius,
in a contrast with 3-D modeling of the radiation transfer in 
the stellar and dust disks.
Since we analyze the vertical profiles separately from each other,
this approach is insensitive to possible bending of the 
galactic disks. Note that if the resulting scaleheight $z_0(r)$ 
demonstrates a systematic growth with distance to the center, 
it suggests that the real radial gradient of the disk scaleheight may 
be a few times greater due to the projection effects. 

We fit $z_0(r)$ with a linear function of radius $r$ and 
find the radial gradients of
$z_0$ (expressed in units of $z_0$) per scalelength, see Figure~2.
Different lines in Figure~2 show histograms of distribution of the
gradient in the J, H, and K$_s$ bands in our whole sample.
The median gradient $(dz_0/dr)\cdot(h/z_0)$ in the $K_s$ is 
essentially zero (-0.01), as well as in the H (0.) and J (0.01) bands.

Note that some galaxies in histogram in Figure~2 indeed show noticeable  
radial gradients of the scaleheight. In combination with projection
effects, it suggests significant gradients $(dz_0/dr)\cdot(h/z_0)$ 
in a small fraction of our galaxies.

\subsection{Reliability of the Structural Parameters}
\label{im_sim}

The dust disks embedded into the stellar components of the galaxies 
may significantly distort the structural parameters of the stellar disks
estimated even from the NIR images. 
We check 
reliability of the estimated parameters via the following simulations. 
A set of artificial images is created using equation
(\ref{vol_disk}) and typical parameters $h$ and $z_0$. The surface brightness
$\mu_0$ is replaced by the luminosity surface density $I_0$ in our
simulations. 
Each stellar disk contains an embedded dust disk inside with scalelength
of 1.5$h$ and scaleheight of 1/3 $z_0$ \citep[according to][]{bianchi07}. 
The face-on
opacity of the dust disk corresponds to $A_V$ = 1 mag. This value 
should be close to the upper limit for the face-on extinction in spiral 
galaxies \citep{kuchinski98,xi99,bianchi07}, 
Finally, the model disks are randomly inclined by 85-90 degrees and 
their images (i.e. 2-D projections to the plane of sky) are obtained 
via integration along the line-of-sight.
The images are then convolved with the instrumental profile typical for the
observations. Then the poisson and gaussian noise patterns are added to
create sets of typically thirty artificial images with the same 
parameters but with different random noise pattern. The images are analyzed 
by the same way as the real galaxies; 
the mean values of $h$, $z_0$, and $I_0$ and their r.m.s. are estimated 
and then compared with the input parameters.  
Figures~3 and 4 give examples how well
the structural parameters can be extracted from the artificial 
edge-on disks made with stars and dust should they be observed in the
J and K$_s$ NIR bands. Although the scalelength and the central surface
brightness can be recovered
with a fairly good accuracy from both J and K$_s$ bands, the scaleheight 
in the J band looks systematically thicker due to the dust absorption
effects only. The K$_s$ scaleheight in this case is found to be 
about 15\% greater than the model input  scaleheight.

We check how the chosen values of the model parameters affect our conclusions.
The amount of the dust opacity is responsible for increasing the 
estimated scale height
and makes it thicker by 10-15\% in the K$_s$ 
if $A_V$ is of the order of 1 mag. Much higher
values of the central dust opacity, more than 3 mag, make stellar disks
look about twice as thick even in the K$_s$ band. At the same time, 
the disk scalelength shows relatively small variations in this case. 
Variations of the dust scalelength between $h$ and 2$h$ do not change 
the estimated structural parameters significantly. The dust disk thickness
affect the estimated stellar disk thickness in the most power when it is
about a half of $z_0$. Once the dust disk is getting comparable in thickness
with the stellar one, it affects mostly the stellar disk surface brightness,
and not $z_0$. 

The inclination of the stellar disk to the line-of-sight 
is an important parameter and it increases the stellar disk thickness 
estimated in our modeling by more than 15\%, if the inclination is 
getting to be greater than 5 degrees. On the other hand, 
the relatively small inclination has almost no effect on the recovered 
stellar disk scalelength
and central surface brightness. More than 5 degrees inclination in 
galactic disks would create problem
with estimation of $z_0$ \citep[see also ][]{degrijs97}, but such 
galaxies should fail the RFGC catalog criteria and we don't 
expect to have many of highly inclined disks in our sample. 
If the size of the modeling disk is so small that its true scaleheight 
is much less than the seeing, the resulting $z_0$ gets independent of
the original disk scaleheight. Our selection of objects for the analysis 
by size of their major axis prevents such situation.
Given all other parameters fixed, the recovering
scalelength, scaleheight, and surface brightness are independent of the
numerical value of $h$, $z_0$, and $I_0$ unless $z_0$ is getting smaller than
a pixel in the model images.

In general, our modeling suggests that we can estimate the structural
parameters with of order of ten percent precision using our method. 
At the same
time, the sensitivity to the variations of the scaleheight within the same 
galaxy with different amount of extinction (like in the case of
observations in different NIR bands) should be by the order of magnitude 
better. 

\section{The Basic Structural Parameters in Different NIR Bands}
\label{three_band}

As we can see from our observational results shown in Figure~1,
the stellar disks look thicker 
in the J band and thinner in K$_s$ for the same object, in agreement
with our modeling from \S~\ref{im_sim}.
Two factors may be responsible for such a difference in real galaxies:
the vertical gradient of stellar population properties and the dust 
attenuation. Let's consider the effects of the radial and vertical 
gradients of stellar population in galactic disks.
The radial gradients of metallicity in the thin disks 
are of the order of -0.07 dex/kpc for the late-type galaxies
\citep{vanZee98}, and even less in our 
Galaxy \citep{rolleston00,daflon_cunha04,carlos06}.
According to the population synthesis models by \citet{bc03},
this creates gradients in (J-K$_s$) color 
of the order of -0.015 mag/kpc (i.e. the periphery is
bluer). Hence, the scalelength in J should appear 4 \% longer 
than that in K for a typical scalelength values of 3 kpc.
A radial age gradient in combination with the metallicity change
may rise this scalelength change to 5\%.
As we see, these numbers are within our accuracy of estimation of the 
scalelength.
The vertical gradients of metallicity and age, and hence 
contribution of the stellar population effects to the vertical color 
variations within the thin stellar disk, are usually 
small \citep{tadross03,seth05,carlos06}. Even if we assume the large 
values for the vertical gradient of [Fe/H] as in \citet{deGrijs00}, 
-0.2 dex/kpc, this would create less than 1\% longer scaleheights
in the J than that in the K$_s$ due to the stellar population gradients. 
The amplitude of the changes of scalelengths and scaleheights between the 
H and K$_s$ bands is roughly a half of that between the J and K$_s$. 

We conclude that the scaleheights of the pure thin stellar disk (i.e. without
dust) estimated from the J, H, and K$_s$ images should appear the same.
All significant scaleheight variations have to be addressed to the
reddening effects. This is in agreement with conclusions 
by \citet{dalcanton02}.
Note that in some cases of extraordinary objects like a superthin galaxy
UGC~7321 \citep{matthews01}, the reddening alone cannot explain the 
observed optical color (B-R), (R-H)
variations in the vertical direction and vertical gradients of 
metallicity and age are required for the stellar population.
At the same time, \citet{matthews01} note that
the superthin's H-K color gradients might be explained by the
wavelength-dependent dust absorption alone.

\section{Structural Parameters of the Thin Stellar Disks Corrected for
Extinction}

Following \citet{kylafis87}, we consider a vertical 
luminosity profile of an edge-on image of a stellar disk
with the volume brightness $\rho_L(z) = \rho_{L0} \, exp(-z/z_0)$. 
Embedded co-planar dust disk with its scaleheight $z_d < z_0$
absorbs and scatters the stellar light in dependence of the dust extinction 
coefficient $\kappa_{\lambda} = \kappa_{\lambda}0 \, exp(-z/z_d)$. 
For simplicity, we neglect the dust scatter appealing 
to \citet{bianchi07} conclusion that 
the structural parameters of stellar disks in 3-D modeling 
of the light distribution with or without scattering are not 
significantly different. 
As in \citet{kylafis87}, the brightness distribution 
along the minor photometric axis (z) in such a disk is 
$I(z) = I_0(z) \frac{1 - \exp(-\tau(z))}{\tau(z)}$, where $I_0(z)$ is the 
brightness profile in the case of zero extinction, and
$\tau$ is the optical depth across the whole edge-on disk, 
which also depends on the vertical distance
over the galactic plane $z$. 
For small $\tau$, $I(z) \approx I_0 (1 - 0.5\, \tau(z))$. 
The observed scaleheight $z_e$ of the stellar disk with added extinction
is connected with its projected brightness along the selected profile as
$I(z) = I_{ce} exp(-z/z_e)$. Given $I_0(z) = I_c \, exp(-z/z_s)$, we
find that $1/z_e \approx 1/z_s (1 - \tau(0))$. 
Therefore, estimating the thickness
of the stellar disk with dust from its edge-on profiles we expect to find 
\begin{equation}
\label{z_e}
z_{e\lambda} = z_s (1 + C_{\lambda}\tau_{V}), 
\end{equation}
where $z_{e\lambda}$ is the $z_e$ estimated from the disk's brightness 
profile at a certain photometric band, 
and the optical depth in the plane of the disk in certain photometric
band is $\tau_{\lambda} = C_{\lambda}\tau_{V}$. We assume that $C_{\lambda}$
equals to 0.276, 0.176, and 0.112 in the J,H, and K$_s$ bands, respectively
\citep{schlegel98}. Since the V-band optical depth $\tau_{V}$ and $z_s$ 
are fixed for the same disk, we use linear regression and fit a line
to three data points (for J,H, and K$_s$ values) 
at $C_{\lambda}$ - $z_{e\lambda}$ diagram to find $z_s$.

For the convenience of comparison of the vertical scales, 
we normalize the J and H scaleheights in each galaxy
by their K$_s$-band scaleheight. The upper panel in Figure~5
compares the normalized scaleheights. Each line in the panel 
connects three thicknesses for one galaxy. 
The tendency to observe the stellar disks thicker in the J than in
the K$_s$ is well seen.
The histogram in the lower panel in Figure~5
shows the distributions of all normalized $z_0$ in three NIR colors.
The vertical thick line designates all scaleheights in the $K_s$ because of the
normalization. The mean ratio of $z_0$
in the J and H to that in K$_s$ is 1.16 and 1.08, respectively. 

We calculate the scaleheight $z_s$ corrected for the extinction 
(or extinction-free hereafter) using equation (\ref{z_e}), which
gives $z_s \le z_0(K_s)$. The distribution of $z_s$ normalized
by the $z_0$(K$_s$) is also shown in the lower panel in Figure~5. 
The median $z_s$/$z_0$(K$_s$)
is about 0.89, i.e. the thin stellar disks in our galaxies are
11\% thinner than it can be estimated from the analysis of the vertical
profiles from the K$_s$-band images. The extinction-free scaleheights are
given in Table~1. 

We observe no systematic variations in the
scalelength between the NIR bands, as it can be seen in Figure~6, which is
the same as Figure~5 but drawn for the scalelengths.
This is not a surprise because our simulations shown in Figures~3 and 4 
suggest rather small systematic variations of the radial scalelength among 
the NIR bands, and
proximity of the $h$ estimated from the J, H, and K$_s$ bands to its
extinction-corrected value. 
Using the scalelength $h$ from the K$_s$ band and extinction corrected 
scaleheights 
we find that the mean radial-to-vertical scale ratio in the thin 
extinction-free stellar disks of galaxies in our sample is 5.6.

\section{Thickness of the Stellar Disks and Mass of the Dark Halos}

The thickness of the stellar disk is sensitive to the gravitational
potential of the disk + bulge + halo system in the vertical direction
\citep[e.g. ][]{bahcall84}. Therefore, the relative thickness of the stellar
disk $z_0/h$ may provide some information about the spherical-to-disk mass 
ratio. As a first step, we consider a very simplified model of a galaxy that
consists of a stellar exponential disk embedded into a spherical dark halo.

The total mass of the exponential stellar disk with the central surface
density $\Sigma_0$ is 
\begin{equation}
\label{z0h_1}
M_d = 2 \pi \Sigma_0 h^2 .
\end{equation}
The total mass of all components of the galaxy is 
\begin{equation}
\label{z0h_2}   
M_t \sim V^2 h ,
\end{equation}
where $V$ is the disk's rotational velocity. 
Following \citet{zasov91} and \citet{kregel05},
we assume that the disk thickness depends on the local surface density
$\Sigma$ and vertical velocity dispersion $\sigma_z$:
\begin{equation}
\label{z02}
z_0 \sim \sigma_z^2 / G \Sigma
\end{equation}
and the ratio of
the vertical-to-radial velocity dispersions is $\sigma_z/\sigma_r = const$.
This is a rather coarse but fair approximation, 
see \citet{polyachenko77,bottema93,gerssen00,kregel_k05}.

We incorporate an assumption of the marginal stability of the stellar
disk, i.e. 
\begin{equation}
\label{z0h_3}
\sigma_r = Q \cdot 3.36 \Sigma/\ae, 
\end{equation}
where the Toomre parameter $Q = const$ \citep{toomre64}. 
Here $\ae$ designates the epicyclic frequency, and $\ae \sim V/h$ for the 
flat rotation curves \citep[see][p.121]{gal_dyn}.
Substituting $\sigma_z$ and $\Sigma$ from equations (\ref{z0h_1}), 
(\ref{z0h_2}), and (\ref{z0h_3})
into (\ref{z02}), we find $z_0 \sim Q^2 \Sigma^2 / (\ae^2 \, G \, \Sigma)$
$\sim \Sigma \, h^2 / (V^2 \, G)$ $ \sim (M_d/M_t) h$.
Therefore,
\begin{equation}
\label{zh}
z_0/h \sim (M_d/M_t) .
\end{equation}

Numerical N-body simulations \citep{zasov91,mikhailova00} in which the 
spherical subsystem was introduced as a fixed potential
confirm the equation (\ref{zh}). Figure~7 is adopted
from \citet{mikhailova00} with additions from later simulations by 
Khoperskov (private communication). Figure~7 shows that the stellar disk
thickness $z_0/h$  is sensitive to the relative mass of the spherical
component $M_s$ normalized by the $M_d$.
Both $M_s$ and $M_d$  are determined within the limits of the stellar disk 
(four scalelengths in the simulations). Circles in Figure~7
correspond to different numerical models with various parameters of the
disk and dark halo subsystems, as well as with different bulges. 
The upper curve corresponds to the bulgeless models, whereas the lower one
is for the models with some bulge contribution. 

Contemporary and more realistic N-body simulations include "live" halos 
that respond to the gravitational potential of disks. 
The relation $z_0/h$ versus $M_s / M_d$ should look very similar
in those simulations as in Figure~7
for the galaxies in which the spherical subsystems dominate by mass
(i.e. for large $M_s/M_d$). 
For the spherical subsystems with masses $M_s \la M_d$,
Figure~7 illustrates an extreme case, and its top left part (large
$z_0/h$ and small $M_s/M_d$) demonstrates the lower limit for mass of 
the spherical subsystem given disk thickness $z_0/h$. Preliminary 
results from N-body simulations (Khoperskov, private communication)
with live halo suggest not essential
corrections to $M_s/M_d$ inferred from $z_0/h$ according to 
Figure~7 (of the order of 15\%).
We encourage reader to assume our further estimations of halo-to-disk 
mass ratio made from the disk thickness as lower limits for the 
real $M_s/M_d$.

Following Figure~7, we calculate the distribution of the dark halo mass
$M_{halo}$ expressed in the units of the stellar disk mass for our galaxies.
The RFGC members are "flat" galaxies that
are not expected to harbor large bulges, so we assume here that $M_{halo} 
= M_s$. We find the following relation for bulgeless
galaxies from Figure~7: $M_{halo}/M_d ~=~ -0.9419 ~+~  0.3737 (h/z_0)$.
According to this formula, the median value 
of $M_{halo}/M_d$ is about 1.3 within our sample. 
The histogram of distribution of $M_{halo}/M_d$ is shown in Figure~8.
No any correlations between $M_{halo}/M_d$ and absolute magnitude, maximum of
rotational velocity, or radial gradient of the scaleheight is found for our
galaxies. A tendency to observe higher $M_{halo}/M_d$ in the 
objects with larger scalelength $h$ is found, although this relation 
shows large uncertainty.

Following the same basic assumptions, 
we can include the central surface density $\Sigma_0$ into consideration.
As it follows from our definition of the disk and total galaxy mass,
$M_d / M_t \sim \Sigma_0 h^2 / V^2 h = \Sigma_0 h / V^2$. 
We perform a linear fit between $\log V$ and $\log h$ for our galaxies 
with known $V$ and found that $h \sim V^{1.42}$. 
It seems more correct to utilize non 
edge-on galaxies to study such a correlation,
and \citet{kormendy90} and \citet{walker99} suggest $h \sim V^{1.4}$ 
and $h \sim V^{1.5}$, respectively, for arbitrary inclined disks. 
If the latter relation is accepted,
it gives $M_d / M_t \sim \Sigma_0 / V^{0.5}$, and in
combination with equation (\ref{zh}) we come to the following relation
between the disk thickness and its central surface density 
\begin{equation}
\label{zh_s0}
z_0/h \sim  \Sigma_0 / h^{1/3}. 
\end{equation}
The upper panel in Figure~9 shows the relation between $h/z_s$
(i.e. extinction-corrected inverse thickness of the disks)
and the central surface brightness in the $K_s$ band $\mu_{0}(K)$. Here 
$\mu_{0}(K)$ was corrected for the extinction according to formula 
(\ref{z_e}). The correction is rather small, and its median value for 
our sample of galaxies is about 0.1 mag/arcsec$^2$.

The lower panel 
in Figure~9 demonstrates the plot of $h/z_s$ versus the central
surface density estimated from observations as
$2.30(M/L) 2.512^{24 - \mu_0}$.
The numerical coefficient in the formula
corresponds to the K$_s$-band solar absolute magnitude 3.33 mag.
We assume the constant (M/L$_{Ks}$) = 1 here.
The trend in Figure~9 is in agreement with formula (\ref{zh_s0}). 
The very big scatter of points in Figure~9 can be attributed to the 
relatively low accuracy in the $\mu_0$, $h$, and $z_s$, 
and even more uncertainty is due to the non-constant value of the 
mass-to-light ratio (M/L).
The scatter of this relation can be lowered dramatically if only
large ($h > 4$ kpc) galaxies with the scaleheight gradient
in the $K_s$ $(dz_0/z_0)/(dr/h) < 0.05$ are considered, see Figure~10. 
In this case the inverse disk thickness ($h/z_s$) depends on $\mu_0(K)$ as
\begin{equation}
   h/z_0 ~=~  47.557  - 23.069\,x + 3.016\,x^2  ~~,
\end{equation}
where $x = log(2.3\cdot2.512^{24- \mu_0(K)})$ = $0.4*(25.11 - \mu_0(K))$.
The solid curve in Figure~10 represents this equation. 
Apparently, the large spiral galaxies whose stellar disks are not flared 
can be better described by the toy-model developed above. Establishing the 
correlation between the $\mu_0$ and $h/z_s$ may be useful for estimating
the disk thickness and $M_{halo}/M_d$ in disky galaxies with arbitrary 
inclination.

Note that our model is oversimplified, and attempts to estimate the 
$M_{halo}/M_d$ ratio only. Should we need to know the values of $M_{halo}$
or $M_d$ taken apart, or structural parameters of the dark halo, 
the rotation curve modeling has to be incorporated. 
Such additional parameter as the disk thickness helps to decrease the
ambiguity during the rotation curves decomposition.
A special attention should be given to correspondence between the 
photometrically and dynamically inferred parameters of galactic subsystems. 
An example of N-body simulations of an early-type
disk galaxy \citep{abadi03} demonstrates that photometry may fail to trace
the dynamically distinctive subsystems in the central parts of
bulge-harboring galaxies. In our consideration we avoid central parts
of galaxies, and using the RFGC objects prevents us from considering
bulge-dominated stellar systems. 

The weakness of the simplified assumptions in the modeling described above
encourages us to apply a more realistic simulations of the
stellar disk vertical structure. Since it requires extensive numerical
simulations and better quality infrared data, and will gain from the
incorporation of rotation curves decomposition, we defer such more
complex consideration to a further paper.

\section{Summary and Conclusions}

We study how the dust absorption affects the basic structural 
parameters (disk central surface brightness, vertical and radial scales)
of stellar disks in spiral galaxies estimated from the NIR J,H, and K$_s$
images. Using 2MASS data, we compare the structural parameters 
estimated from different NIR bands focusing on the scaleheight of thin 
stellar disks. The stellar disks look thinner in the 2MASS K$_s$ band in the
comparison with the H and J. We employ this fact to figure out the
extinction-corrected scaleheight $z_0$ of the thin stellar disk. 

Using 139 relatively large galaxies selected from the 2MASS catalog, 
we find that the 
mean vertical scaleheight ($z_0$) ratios are 1.16:1.08:1.00:0.89 in
the J:H:K$_s$:extinction-free bands, respectively. This means that the
radial scalelength estimated from the $K_s$ images is very close to the
extinction-corrected one, whereas the scaleheight is overestimated
by 11\% on average. The mean extinction correction for the face-on central
surface brightness is only about 0.1 mag/arcsec$^2$ in the K$_s$.

The median radial-to-vertical scale ratio for our sample is about 6.
Using a relation between the stellar disk thickness and the
halo-to-disk mass ratio obtained from a simplified model, 
we estimate the dark-to-luminous mass ratio 
in our galaxies within the limits of their optical disks. 
Its mean value is about 1.3. 
A relation between the stellar disk thickness and its
central surface brightness is observed for the galaxies from our sample, 
although it shows significant scatter and is affected by uncertainty in 
additional parameters. This relation is more prominent for 
large ($h > 4$ kpc) galaxies from our sample and can be utilized for 
coarse estimating the disk thickness and relative mass of the
dark halo in large spiral galaxies arbitrarily inclined to the line-of-sight.

We find also that the scaleheight does not change systematically along the 
radius in most of our galaxies. Only a small fraction of galaxies
demonstrates a noticeable radial growth of the scaleheight.

\acknowledgments

This project was partly supported by grant RFBR-07-02-00792.
This publication makes use of data products from the Two Micron All Sky
Survey, which is a joint project of the University of Massachusetts and the
Infrared Processing and Analysis Center/California Institute of Technology,
funded by the National Aeronautics and Space Administration and the National
Science Foundation. Computational equipment for DB was provided by AAS
under a Small Research Grant. We appreciate recommendations from anonymous 
referee that improved the paper. We acknowledge the usage of the HyperLeda
database (http://leda.univ-lyon1.fr).

\clearpage

\begin{table}
\begin{center}
\caption{The Structural Parameters of the Galaxies}
\begin{tabular}{lcccccccccc}
\tableline\tableline
Name      & $h$(K) & d$h$ & z0(K)&  d$z_0$ &$\mu_0$(K)&d$\mu_0$&$z_0$(cor.)&$z_0$(cor)/$h$&$\mu_0$(K,cor)&$\frac{M_{dark}}{M_{lum}}$\\
          & kpc  &  kpc  &  kpc  &  kpc&mag/$\Box$''&mag/$\Box$''&kpc &-- &mag/$\Box$''& -- \\
\tableline
RFGC0095 &  0.72 &  0.06  & 0.31 &  0.08 & 17.69 &  0.26 &  0.27 & 0.375 & 17.61  & 0.06\\
RFGC0099 &  4.35 &  0.06  & 0.95 &  0.17 & 17.51 &  0.18 &  0.83 & 0.191 & 17.41  & 1.02\\
RFGC0102 &  1.76 &  0.03  & 0.48 &  0.11 & 17.49 &  0.20 &  0.46 & 0.260 & 17.46  & 0.49\\
RFGC0124 &  8.15 &  0.24  & 1.36 &  0.19 & 17.94 &  0.18 &  1.12 & 0.137 & 17.81  & 1.79\\
RFGC0152 &  5.33 &  0.06  & 0.88 &  0.14 & 17.73 &  0.21 &  0.76 & 0.143 & 17.66  & 1.67\\
\multicolumn{6}{l}{...}\\
\multicolumn{6}{l}{The full table is available on-line}\\
\tableline
\end{tabular}
\end{center} 
\vbox{
The table shows RFGC name, scalelength $h$ in kpc and its uncertainty in the
K$_s$ band, scaleheight $z_0$ in kpc and its uncertainty in the K$_s$,
central surface brightness $\mu_0$ in the K$_s$, extinction-corrected
scaleheight $z_0$(cor.) in kpc, extinction-corrected relative thickness of
the stellar disk $z_0(cor.)/h$, extinction-corrected central surface
brightness $\mu_0(cor.)$ in the K$_s$, and dark-to-luminous mass ratio
$M_{dark}/M_{lum}$.
}
\end{table}

\clearpage
\begin{figure}
\includegraphics[scale=.80]{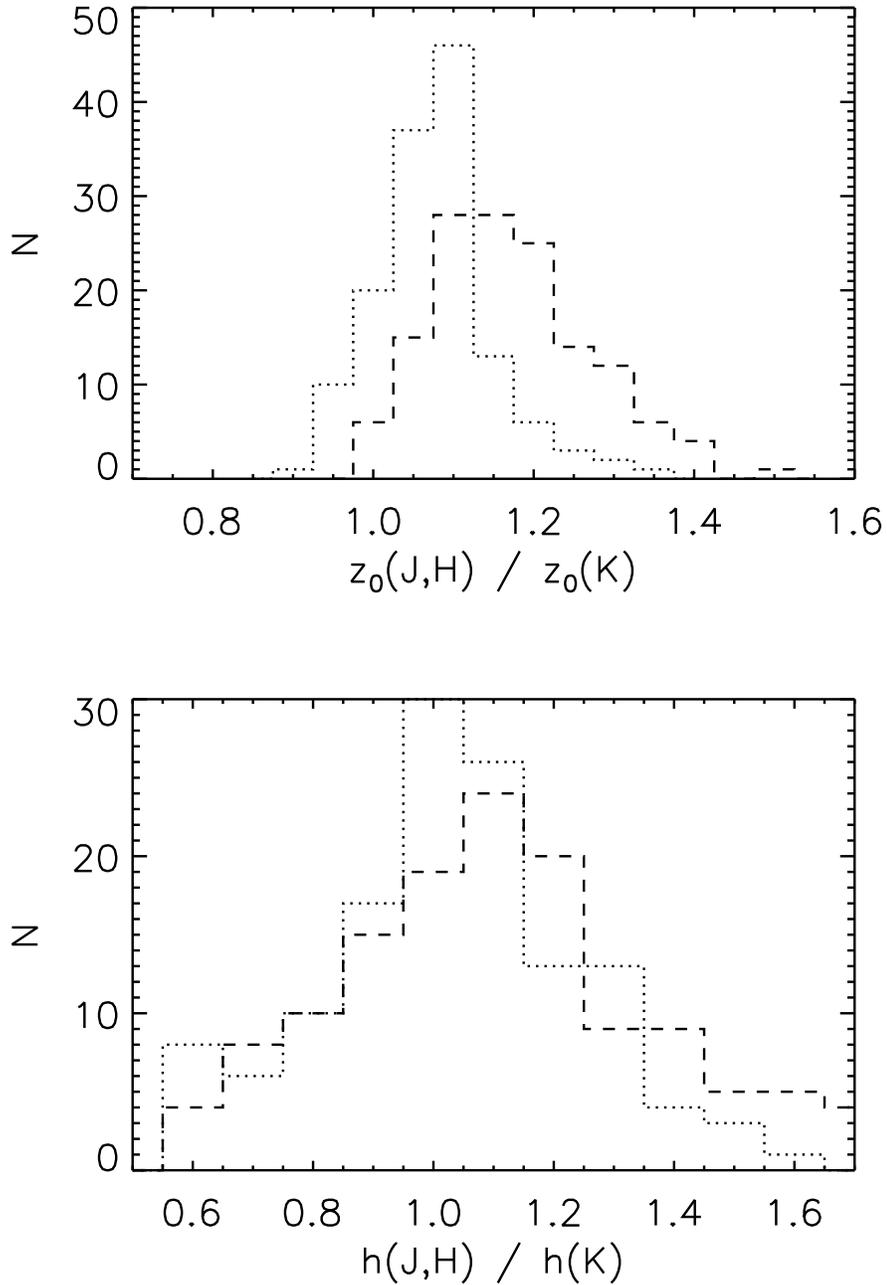}
\caption{
The scaleheights (top panel) and scalelenghts (bottom panel) 
in the J (dashed line) and H (dotted line) bands normalized by those from
the K$_s$ for our 139 galaxies. The K$_s$ band scales in both panels  
equal to unit for all objects.
\label{fig0}}
\end{figure}

\clearpage
\begin{figure}
\includegraphics[scale=.80]{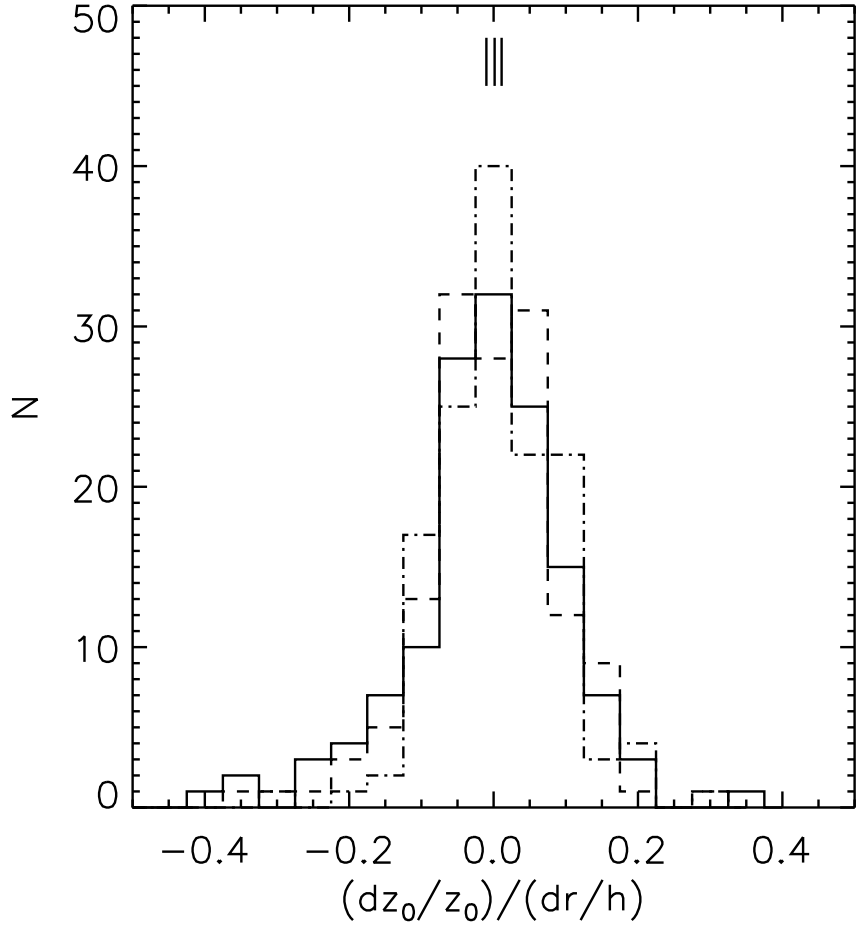}
\caption{
Distribution of the radial gradients of the scaleheight
per one scalelengths $(dz_0/z_0)\, /\, (dr/h)$. The solid, dashed, and
dash-dotted lines designate the $K_s$, H, and J gradients, respectively.
The three short vertical lines above the histogram indicate the median 
values of the gradients in the $K_s$, H, and J bands (from left to right),
that are essentially equal to zero.
\label{fig4}}
\end{figure} 

\clearpage
\begin{figure}
\epsscale{.70}
\plotone{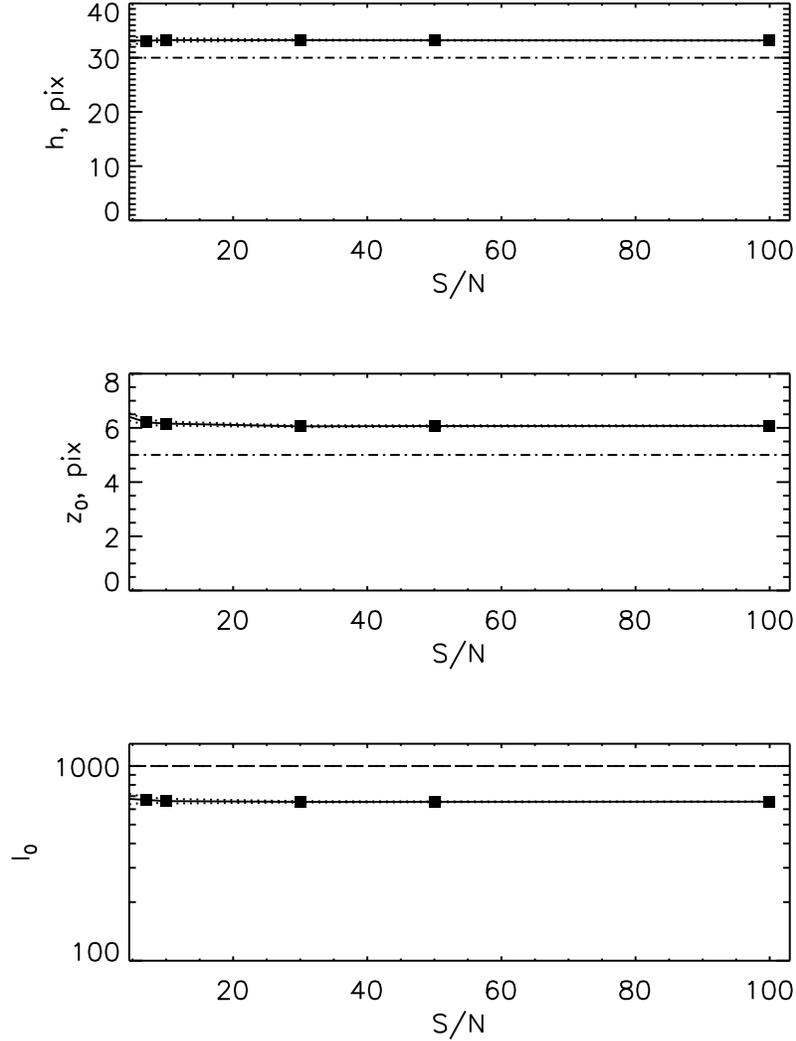}
\caption{The structural parameters of the dusty stellar
disks estimated from sets of artificial images in the J band are shown by 
the solid lines with squares for different signal-to-noise ratio (S/N in the 
center of the disks is used). The dotted lines designate the r.m.s scatter
of the parameters. The dash-dotted horizontal lines denote the input (i.e. 
true) model parameters. The scalelength, scaleheight, and central surface 
luminosity ($I_0$, in arbitrary units) are shown in the upper, middle, and 
lower panels, respectively. 
\label{fig1}}
\end{figure}

\clearpage
\begin{figure}
\includegraphics[scale=.80]{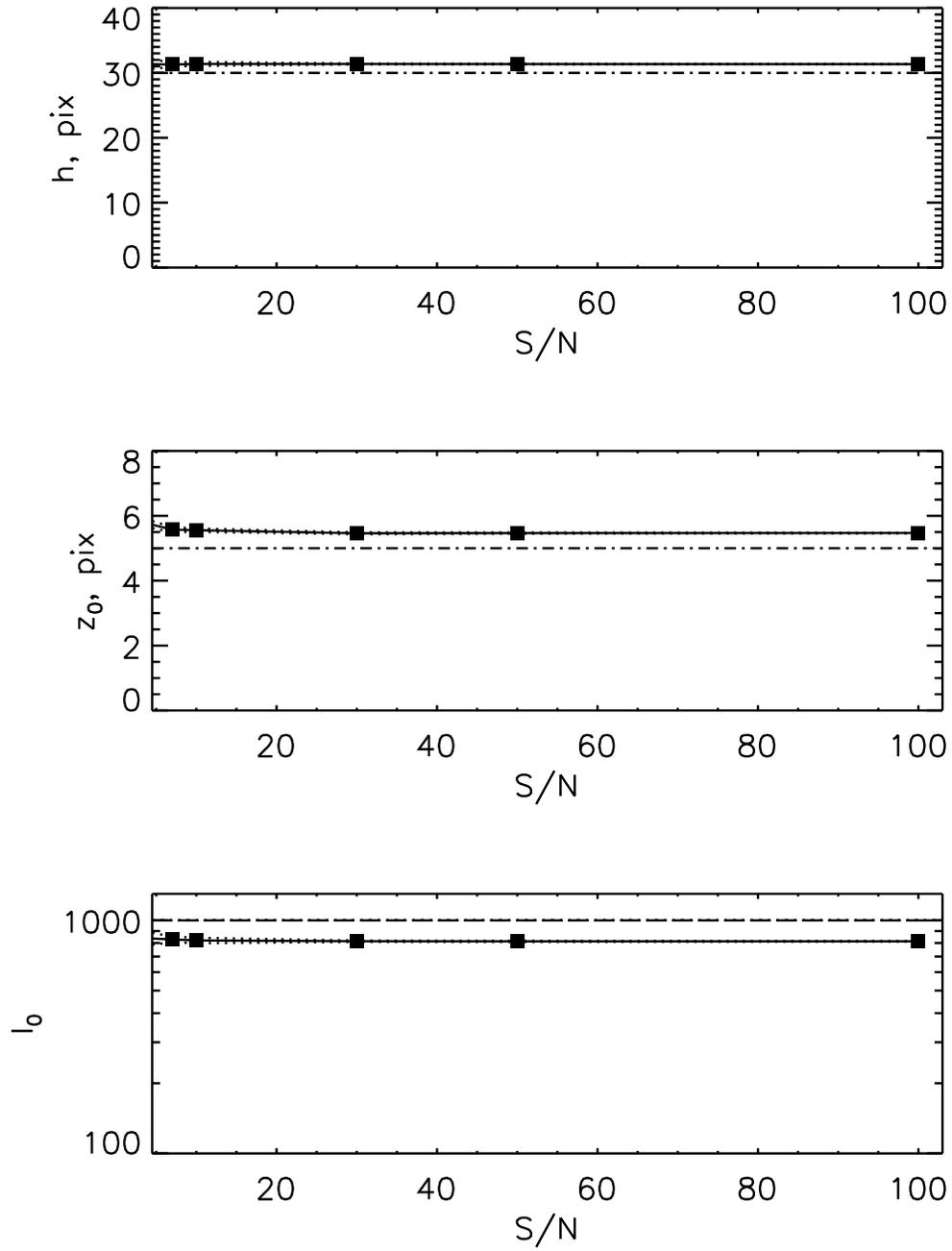}
\caption{The same as in Figure~\ref{fig1} but for the K$_s$ band.
\label{fig2}}
\end{figure}

\clearpage
\begin{figure}
\includegraphics[scale=.80]{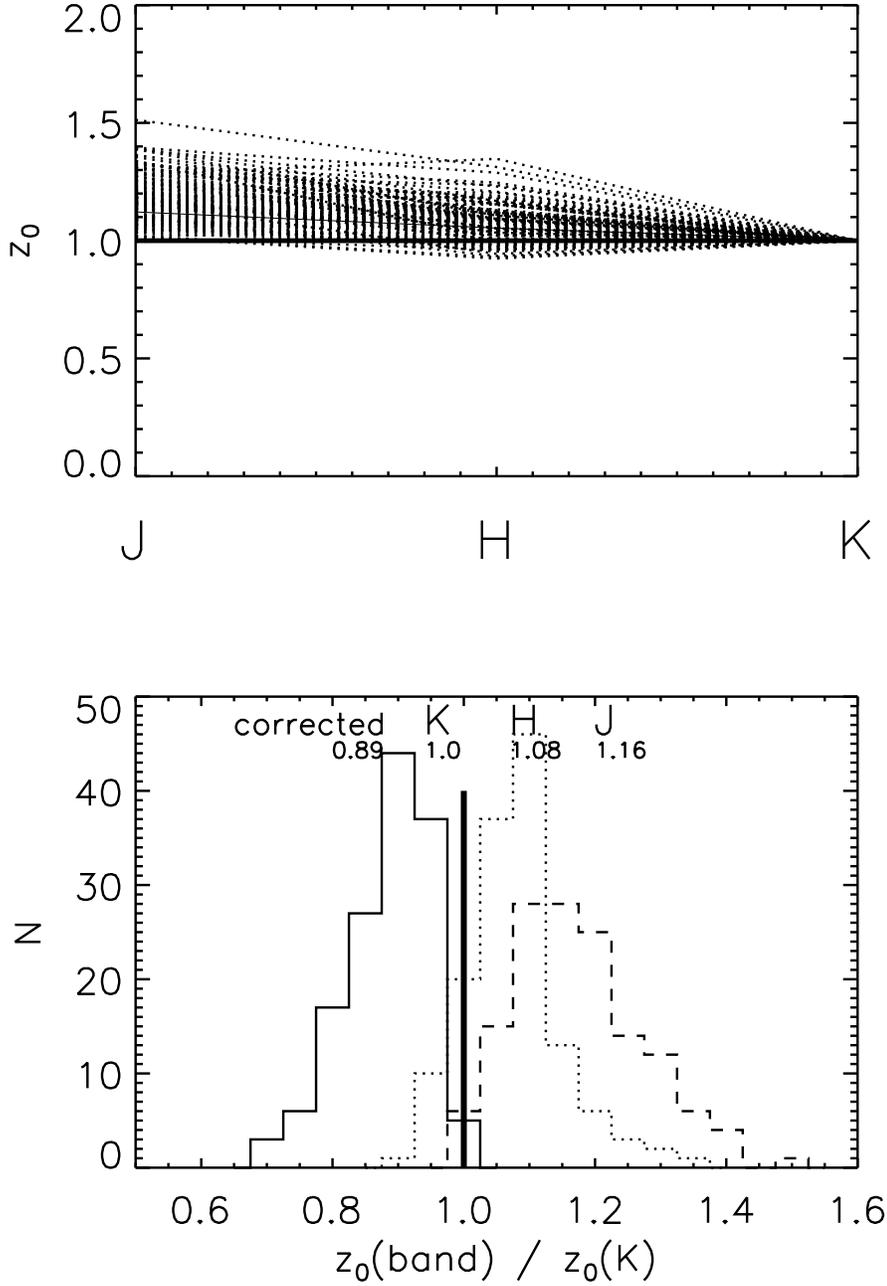}
\caption{Top panel: all scaleheights $z_0$ normalized by that in the K$_s$.
Each a galaxy is shown by a dashed line, which connects the $z_0$
values in the J, H, and K$_s$ (the latter equals to unit for all
objects). The solid line shows the case of equality of all scaleheights.
Bottom panel: histograms of distribution of the scaleheights normalized by 
those in the K$_s$ band. Dashed line corresponds to the J band, dotted - 
to the H band. The solid line designates the extinction-corrected scaleheights. 
The numbers under the letters show the mean values of the scaleheights in
the corresponding bands. The thick vertical line designates formally the
scaleheights in the $K_s$ that are all equal to unity due to the normalization.
\label{fig3}}
\end{figure} 

\clearpage
\begin{figure}
\includegraphics[scale=.80]{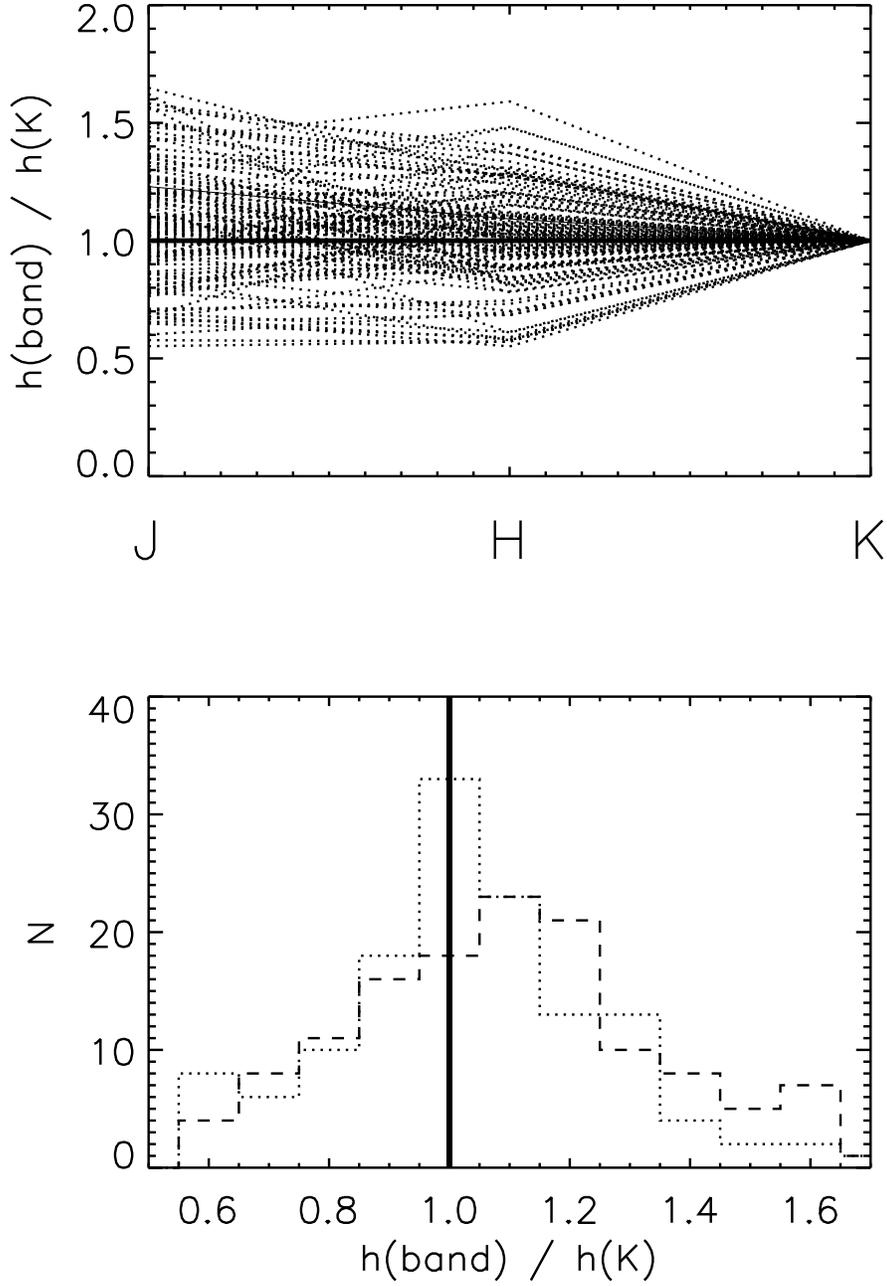}
\caption{
Top panel is the same as in Figure~5 but drawn for the disk scalelengths. 
Bottom panel is similar to the bottom panel in Figure~5 and is plotted for 
the scalelengths. The scalelengths are normalized by
those from the K$_s$ band. Dashed line corresponds to the J band, dotted - 
to the H band. The thick vertical line designates formally the scalelengths
in the $K_s$ that are all equal to unity due to the normalization.
\label{fig3}}
\end{figure}

\clearpage
\begin{figure}
\includegraphics[scale=.80]{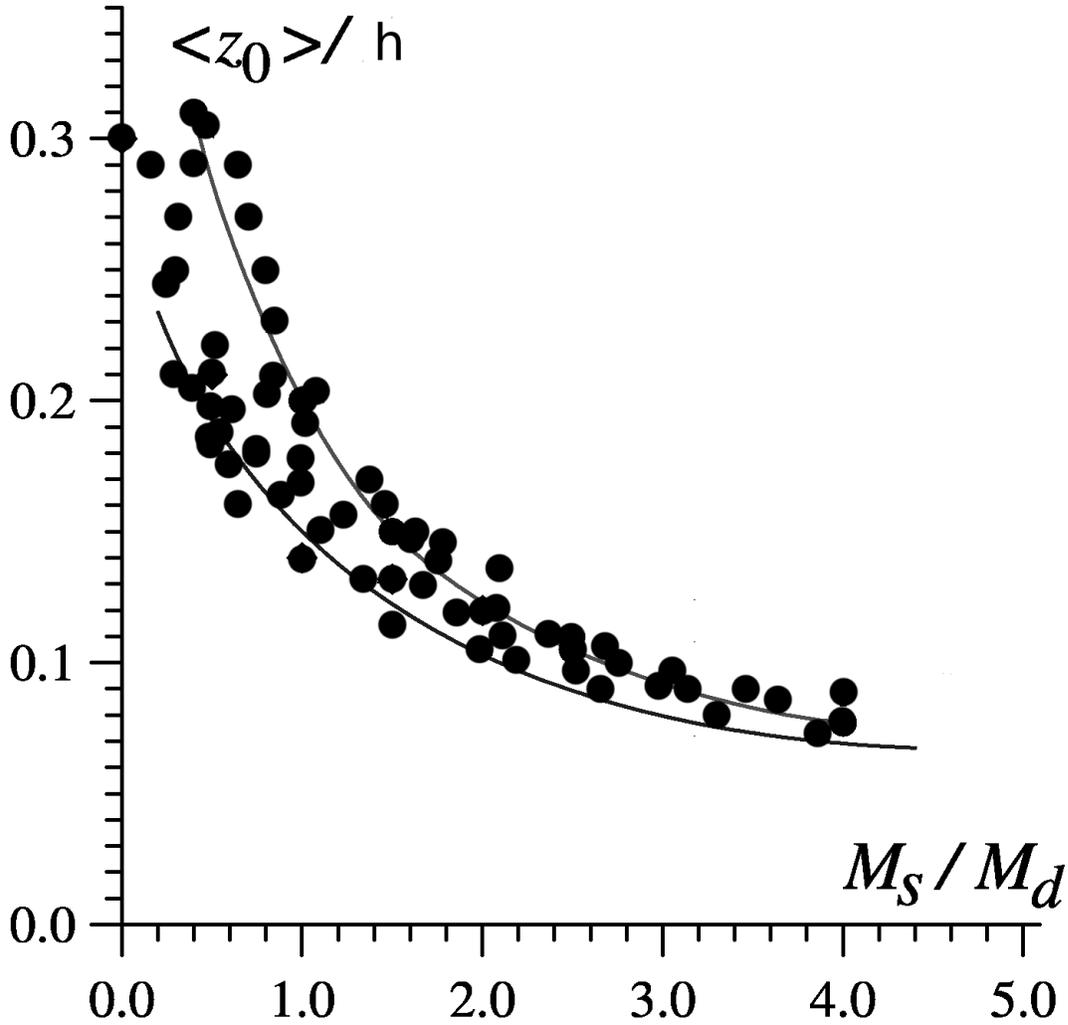}
\caption{The relative disk thickness $z_0/h$ is shown against
the spherical-to-disk mass ratio $M_s/M_d$. Here $M_s$ is the mass of  
the spherical component (halo and bulge, in a general case), and
$M_d$ is the mass of the disk component. 
The $M_s$ and $M_d$ are estimated within the limits of the stellar disk (4$h$).
This figure comes from simplified N-body simulations \citep{mikhailova00} 
and each symbol corresponds to certain numerical model. The models were run
with a variety of parameters for the disk, dark halo, and bulge. 
The upper curve corresponds to the bulgeless models, whereas the lower one
is for the models with some bulge contribution.
\label{fig5}}
\end{figure} 

\clearpage
\begin{figure}
\includegraphics[scale=.80]{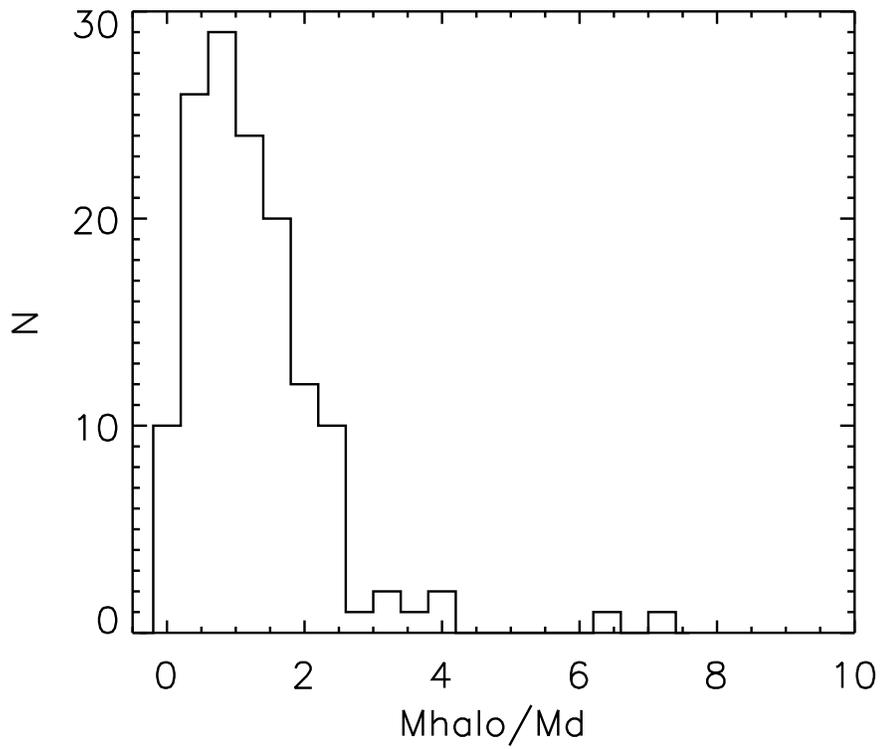}
\caption{The dark halo-to-disk mass ratio for the galaxies from our sample
inferred from the relative disk thickness.
The mass ratio is close to the dark-to-luminous mass ratio since most of 
our galaxies have no prominent bulges. 
\label{fig6}}
\end{figure} 

\clearpage
\begin{figure}
\includegraphics[scale=.80]{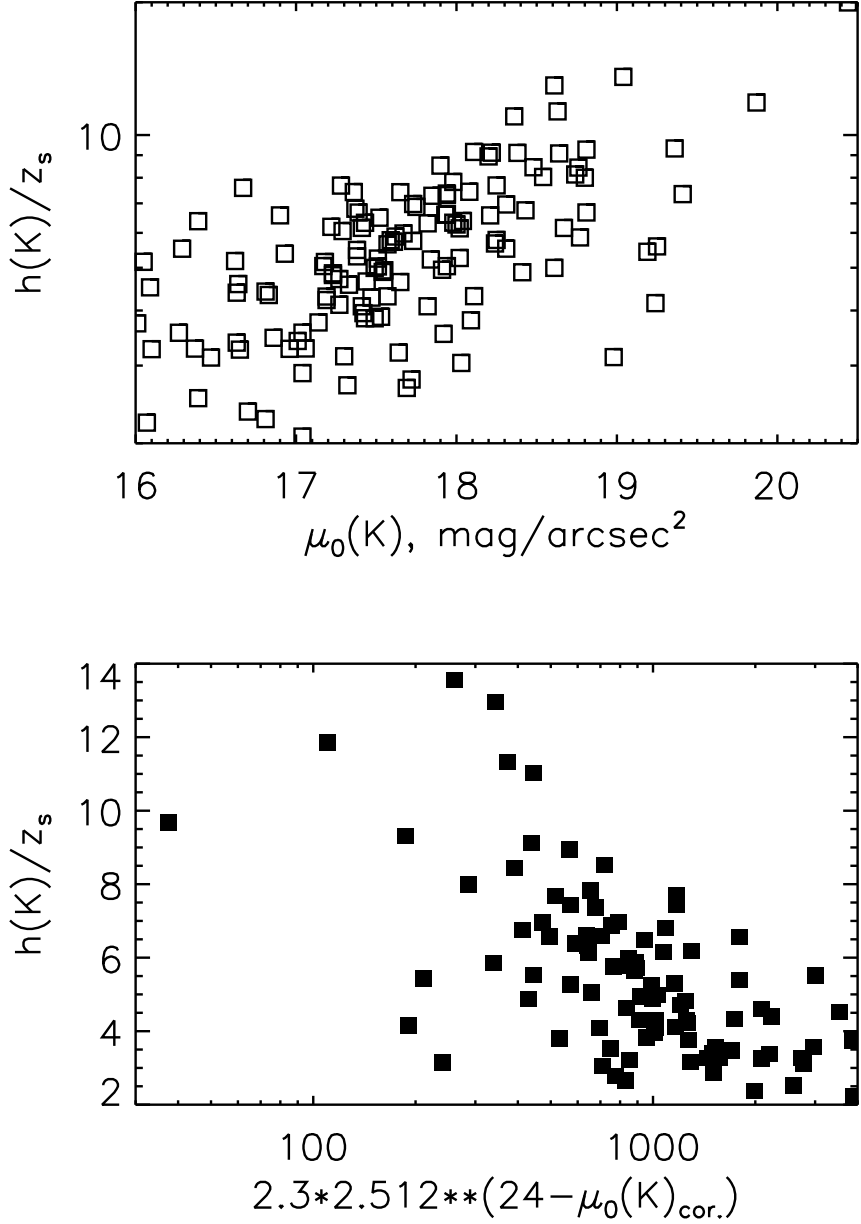}
\caption{Top panel: the inverse disk thickness $h/z_s$ corrected for the
internal extinction versus the central surface brightness in the K$_s$ band. 
Bottom panel: the same relation where $\mu_0$ is replaced with the central
surface density calculated as $2.30 \cdot 2.512^{24 - \mu_0}$, see text.
\label{fig7}}
\end{figure} 

\clearpage
\begin{figure}
\includegraphics[scale=.80]{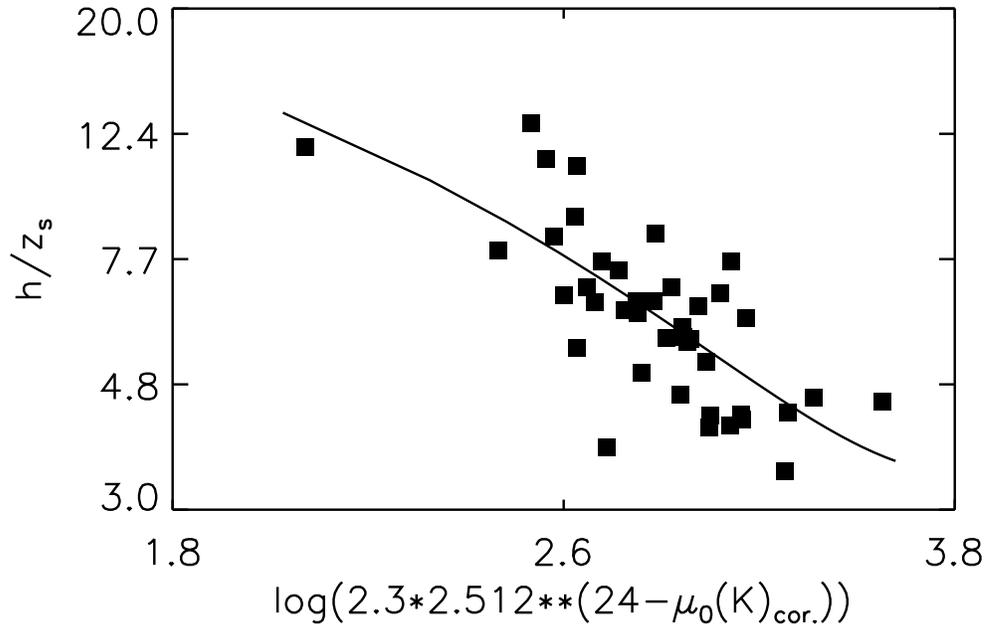}
\caption{The inverse disk thickness $h/z_s$ corrected for the extinction 
versus the central surface density of the stellar disks (see text) 
for the sample of large ($h > 4$ kpc) and non-flared disk galaxies. 
The solid line shows the best-fit to the data (see text).
\label{fig8}}
\end{figure}

\end{document}